\documentclass[journal=jctcce, manuscript=article]{achemso}

\usepackage{chemformula} 
\usepackage[T1]{fontenc} 
 \usepackage[colorlinks=true,linkcolor=blue,urlcolor=blue,citecolor=blue]{hyperref}
\usepackage{array}
\usepackage{lipsum}
\usepackage{bm}
\usepackage{subfigure}
\usepackage{amssymb}
\usepackage{multirow}
\usepackage{tabularx}
\usepackage{amsmath}
\usepackage{braket}
\usepackage{caption}
\usepackage{float}
\usepackage{caption}
\usepackage{enumerate}
\usepackage{ulem}
\usepackage{extarrows}
\usepackage{caption}
\usepackage{mathtools}

\newcommand\underrel[3][]{\mathrel{\mathop{#3}\limits_{%
      \ifx c#1\relax\mathclap{#2}\else#2\fi}}}
      
\newsavebox{\measurebox}

\renewcommand{\vec}[1]{\mathbf{#1}}

\renewcommand{\thefootnote}{*}

\SectionsOn
\SectionNumbersOn

\author{Zhandos A. Moldabekov} 
\email{z.moldabekov@hzdr.de}
\affiliation{Institute of Radiation Physics, Helmholtz-Zentrum Dresden-Rossendorf (HZDR), D-01328 Dresden, Germany}
\alsoaffiliation{Center for Advanced Systems Understanding (CASUS), D-02826 G\"orlitz, Germany}

\author{Sebastian Schwalbe}
\affiliation{Center for Advanced Systems Understanding (CASUS), D-02826 G\"orlitz, Germany}
\alsoaffiliation{Helmholtz-Zentrum Dresden-Rossendorf (HZDR), D-01328 Dresden, Germany}

\author{Uwe Hernandez Acosta}
\affiliation{Center for Advanced Systems Understanding (CASUS), D-02826 G\"orlitz, Germany}
\alsoaffiliation{Helmholtz-Zentrum Dresden-Rossendorf (HZDR), D-01328 Dresden, Germany}

\author{Thomas Gawne}

\affiliation{Center for Advanced Systems Understanding (CASUS), D-02826 G\"orlitz, Germany}
\alsoaffiliation{Helmholtz-Zentrum Dresden-Rossendorf (HZDR), D-01328 Dresden, Germany}

\author{Jan Vorberger}
\affiliation{Helmholtz-Zentrum Dresden-Rossendorf (HZDR), D-01328 Dresden, Germany}

\author{Michele Pavanello}
\affiliation{Department of Physics, Rutgers University, Newark, NJ 07102, USA}

\author{Tobias Dornheim}
\affiliation{Center for Advanced Systems Understanding (CASUS), D-02826 G\"orlitz, Germany}
\alsoaffiliation{Helmholtz-Zentrum Dresden-Rossendorf (HZDR), D-01328 Dresden, Germany}

\title{ Enhancing the Efficiency of Time-Dependent Density Functional Theory Calculations of Dynamic Response Properties}

\abbreviations{IR,NMR,UV}

\begin{document}
\begingroup
\renewcommand\thefootnote{}\footnotetext{Corresponding author: Zhandos A. Moldabekov}%
\addtocounter{footnote}{-1}%
\endgroup

\abstract{
X-ray Thomson scattering (XRTS) constitutes an essential technique for diagnosing material properties under extreme conditions, such as high pressures and intense laser heating. Time-dependent density functional theory (TDDFT) is one of the most accurate available \textit{ab initio} methods for modeling XRTS spectra, as well as a host of other dynamic material properties. However, strong thermal excitations, along with the need to account for variations in temperature and density as well as the finite size of the detector significantly increase the computational cost of TDDFT simulations compared to ambient conditions.
In this work, we present a broadly applicable method for optimizing and enhancing the efficiency of TDDFT calculations. Our approach is based on a one-to-one mapping between the dynamic structure factor and the imaginary time density--density correlation function, which naturally emerges in Feynman’s path integral formulation of quantum many-body theory. Specifically, we combine rigorous convergence tests in the imaginary time domain with a constraints-based attenuation of narrow-band fluctuations to improve the efficiency of TDDFT modeling without the introduction of any significant bias. As a result, we can report a speed-up by up to an order of magnitude, thus substantially reducing the burden of computational cost required for XRTS analysis.
}


%





\section{Introduction}

The advent of high brilliance \cite{Fletcher2015}, high resolution \cite{Descamps2020,Wollenweber_RSI, Gawne_PRB_2024, Gawne_2025}, and high repetition rate \cite{Tschentscher_2017}  X-ray free-electron lasers (XFELs) such as the European XFEL (Germany) \cite{preston2020design, Zastrau_2021, Descamps_RSI_2025}, LCLS (USA) \cite{Ofori_Okai_2018, Fletcher2015}, and SACLA (Japan) \cite{hara2013two} has significantly improved our ability to study the electronic structure properties of materials under both ambient and extreme conditions \cite{ Pascarelli2023}. Notable recent examples of successful XFEL applications in materials science under high pressures and/or high temperatures include the first experimental observations of liquid carbon \cite{Kraus_Nature_2025}, superheated gold crystal lattices beyond the predicted entropy catastrophe threshold \cite{White_Nature_2025}, and phonon hardening in laser-excited gold \cite{Descamps_sciadv}.
Moreover, modern backlighter set-ups~\cite{MacDonald_POP_2022} at high-power laser facilities such as the National Ignition Facility (NIF)~\cite{Moses_NIF} and the OMEGA laser facility~\cite{Glenzer_PRL_2007} (both USA) have, despite their comparably lower spatial and temporal resolution, also become of considerable value for the diagnostics of extreme states of matter, with the recent study of the onset of pressure-driven K-shell delocalization in warm dense beryllium at the NIF~\cite{Tilo_Nature_2023} being a case in point.

From the range of available X-ray diagnostics options, X-ray Thomson scattering (XRTS) has emerged as a primary technique for probing the state of matter under extreme conditions \cite{siegfried_review, Tilo_Nature_2023, Dornheim_T_20222, Bellenbaum_APL_2025}. The resulting need to analyse high-resolution XRTS measurements has highlighted the importance of a true \textit{ab initio} modeling of the probed electron dynamic structure factor (DSF) $S(\mathbf{q},\omega)$~\cite{dornheim2024_Rayleigh}. In addition, high-resolution XRTS data is not only used for diagnostics, but also serves as an important benchmark for theoretical models and to gain insights into the dielectric properties of materials \cite{Sperling_PRL_2015, Sperling_2017, Witte_PRL, Dornheim_NatureCom_2025, Tilo_Nature_2023, Gawne_PRB_2024, Gawne_2025}. 
For example, identifying the applicability range of where common models are accurate is required for the creation of reliable, efficient, and flexible parameterized models of the dynamic collision frequency, electrical conductivity, and other material response properties \cite{Hentschel_POP_2025, Hentschel_POP_2023, Moldabekov_PPNP_2025}. Such models, e.g., constitute critical input for state-of-the-art radiation hydrodynamics simulations \cite{Clark_POP_2015}, which are particularly important for the development of inertial fusion energy (IFE) applications; a field which is currently experiencing significant advancements~\cite{Batani_roadmap,Spears_Science_2025,Betti2023} spearheaded by the recent demonstration of target gain greater than unity in an inertial fusion experiment at NIF\cite{PhysRevLett.132.065102}.

The primary first-principles modeling method to compute the electronic DSF is given by time-dependent density functional theory (TDDFT). Both commonly used flavours---linear-response TDDFT \cite{book_Ullrich, Byun_2020, Moldabekov_PRR_2023, Moldabekov_mre_2025} and real-time TDDFT \cite{Baczewski_PRL_2016, White_Electronic_Structure_2025,Dornheim_review}---utilize a Lorentzian or Gaussian broadening with a finite width $\eta$ to suppress numerical fluctuations in the results for electron dynamic response functions and DSFs. This broadening often obscures important features of the DSF, making them difficult to discern. Theoretically, the correct limit for any response property would be achieved by setting the broadening parameter to $\eta\to 0$. However, in practice, $\eta$ must be set to a finite value to extract sufficiently smooth curves for DSFs. 
For fixed simulation parameters, a decrease in $\eta$ results in the amplification of irregularities, which is characterized by evident ringing in the curves. Consequently, to achieve a smooth curve with lower $\eta$ values, a finer k-point mesh for sampling the first Brillouin zone is required. 
First and foremost, this presents a significant bottleneck for TDDFT as the computational cost increases cubically or even faster with the density of $k$-point mesh, coupled with the already inherently high computational demands of TDDFT. Secondly, the high computational cost of the TDDFT significantly obstructs the convergence tests for the DSF and its related dynamic response properties with respect to the parameter $\eta$. 

While ideal crystal structures exhibit lattice symmetries that help to mitigate these issues \cite{MacDonald_1979}, these problems become much more severe in systems with strong thermal excitations due to a significant increase in the number of partially occupied bands as the temperature rises \cite{Blanchet_POP_2020, Fiedler_PRR_2022}. Moreover, for systems with disordered atomic structures such as disordered alloys, amorphous materials, and shock-compressed materials, techniques that employ crystal symmetries are not applicable. Additionally, for such materials, the computational cost increases dramatically due to the necessity of averaging the TDDFT results over various ionic configurations \cite{JCP_averaging}.
Consequently, the optimal value of the broadening parameter $\eta$ in such computationally demanding cases is usually selected by visually assessing the smoothness of the TDDFT results for the dynamic response property of interest \cite{abinit_tddft, Meghan_AIP_2023, Desjarlais_PRE_2002}; this is clearly a highly unsatisfactory approach for what is supposed to be a true first-principles calculation.

In this work, we develop a novel metric to test the convergence with respect to $\eta$ that is formulated in the imaginary time domain, and combine it with a constraints-based attenuation of narrow-band fluctuations, eliminating the need to continuously increase the density of $k$-points as $\eta$ decreases without the introduction of any significant bias. The cornerstone of our approach is the one-to-one mapping between the DSF $S(\mathbf{q},\omega)$ and the imaginary-time correlation density--density function (ITCF) $F(\mathbf{q},\tau)$---a key quantity both for the interpretation of XRTS experiments~\cite{Dornheim_T_20222,Dornheim_T2_2022,dornheim2024_Rayleigh,Schoerner_PRE_2023,shi2025firstprinciplesanalysiswarmdense,schwalbe2025staticlineardensityresponse,Vorberger_PRX_2023,Bellenbaum_APL_2025,Dornheim_NatureCom_2025} and, e.g., for first-principles quantum Monte Carlo simulations~\cite{boninsegni1996,Dornheim_JCP_ITCF_2021,Saccani_Supersolid_PRL_2012,Dornheim_mre_2023,Dornheim_MRE_2024,Chuna_JCP_2025}.

The presented techniques boost the efficiency of \textit{ab initio} TDDFT calculations of dynamic response properties for both ambient and extreme conditions, thereby substantially reducing the computational effort required for modeling XRTS spectra. In addition, the combination of the $\eta$-convergence test with the constraints-based attenuation of narrow-band fluctuations enables more efficient TDDFT modeling of a plethora of other dielectric properties, such as the dynamic density response function, dielectric function, conductivity, absorption, etc. Therefore, we expect our work to become a new standard tool for the estimation of material properties and for the theoretical support of experimental measurements over a range of conditions and research fields.
We demonstrate our new approach by applying it to the DSF of solid density hydrogen (with $\rho=0.08 ~{\rm g/cc}$) at parameters relevant for experiments with laser heating of cryogenic hydrogen targets \cite{Fletcher_Frontiers_2022, Zastrau} and IFE \cite{Hu_PRL_2010, Bonitz_pop_2024}. Additionally, we consider aluminum (Al) both under ambient conditions and with isochoric heating \cite{Descamps_sciadv, Descamps2020, Moldabekov_prr_2024, Moldabekov_Omega_2024,Sperling_2015}.  


The paper is organized as follows: In Sec. \ref{s:eta_test}, we present a method for the $\eta$-convergence test of TDDFT results in the imaginary time domain. In Sec. \ref{s:constr}, we introduce a method of constraints-based attenuation of narrow-band fluctuations for TDDFT modeling of the DSF and other related dynamic response properties. In Sec. \ref{s:results}, we demonstrate the utility of the developed $\eta$-convergence test in the imaginary time domain combined with the constraints-based attenuation of narrow-band fluctuations by analyzing the DSF of solid density hydrogen and isochorically heated aluminum. We conclude the paper by emphasizing our main findings and discussing potential applications.
Supporting technical information about TDDFT calculations and data post-processing can be found in Appendix~\ref{sec:methods}.


\section{Method for the $\eta$-convergence test in the imaginary time domain}\label{s:eta_test}

Our proposed approach is based on the one-to-one mapping between the DSF $S(\mathbf{q},\omega)$ and the ITCF $F(\mathbf{q},\tau)$, indicating that both contain precisely the same information about dynamic material properties, only in different representations~\cite{Dornheim_prb_2023, Dornheim_mre_2023, Dornheim_PTR_2022, Chuna_JCP_2025}. These two representations are connected through a two-sided Laplace transform:
\begin{eqnarray}\label{eq:analytic_continuatio}
F(\mathbf{q},\tau) = \int_{-\infty}^\infty \textnormal{d}\omega\ S(\mathbf{q},\omega)\ e^{-\tau\omega},
\end{eqnarray}
where $-i\hbar\tau\in -i\hbar[0,\beta]$ is the imaginary time argument in the range defined by the inverse temperature $\beta=1/(k_BT)$, and the DSF at negative frequencies is computed using the detailed balance $S(\mathbf{q},-\omega)=S(\mathbf{q},\omega)\exp(-\beta \omega)$; the latter is equivalent to the symmetry relation of the ITCF $F(\mathbf{q},\tau)=F(\mathbf{q},\beta-\tau)$, which holds universally in thermal equilibrium~\cite{Dornheim_T_20222,Dornheim_T2_2022}. The ITCF naturally emerges in Feynman’s path integral formulation of quantum many-body theory. Recently, it has gained increasing attention after Dornheim et al.~\cite{Dornheim_T_20222, Dornheim_T2_2022} demonstrated that the two-sided Laplace transform allows for the direct determination of the electron temperature, and a variety of other properties~\cite{dornheim2024_Rayleigh,schwalbe2025staticlineardensityresponse,Bellenbaum_APL_2025,Vorberger_PRX_2023,Dornheim_NatureCom_2025}, from XRTS measurements without the need for any modeling.

\begin{figure}[t!]
  \centering
  \begin{tabular}{ c }
    \includegraphics[width=.4\linewidth]{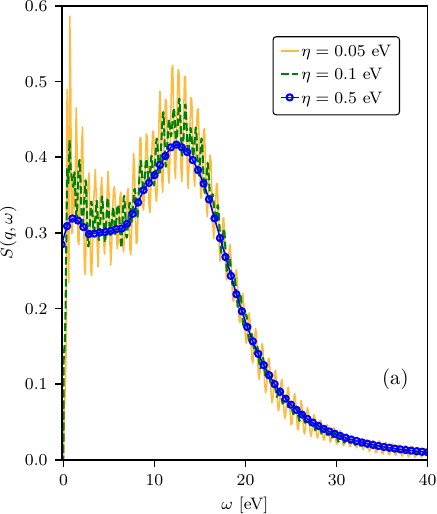}
  \end{tabular}%
    \begin{tabular}{ c }
    \includegraphics[width=.4\linewidth]{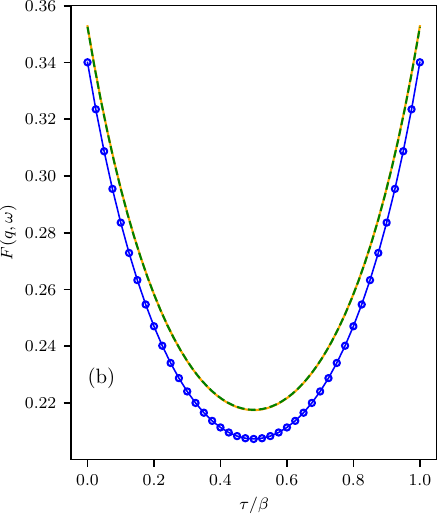}
  \end{tabular}%
  \caption{ TDDFT results for (a) the DSF $S(\mathbf{q},\omega)$ and (b) the ITCF $F(\mathbf{q},\tau)$ of solid density hydrogen with $T=4.8 ~{\rm eV}$ and $\rho=0.08 ~{\rm g/cc}$ at $q=0.946~\textup{~\AA}^{-1}$. Results are shown for different Lorentzian smearing parameters $\eta$. The TDDFT results were averaged using 20 different ion configurations generated using molecular dynamics simulations (see Sec. \ref{s:dft_info} for details). }\label{fig:ilust_h}
\end{figure}

For the current analysis, an important aspect of the Laplace transform is its ability to filter out high-frequency narrow-band
fluctuations \cite{Dornheim_T2_2022}, making it a robust tool for performing the $\eta$-convergence test. To demonstrate this point, we show the results for the DSF and corresponding ITCF computed for solid density hydrogen with temperature $T=4.8 ~{\rm eV}$ and density $\rho=0.08 ~{\rm g/cc}$ at $q=0.946~\textup{~\AA}^{-1}$ and the Lorentzian smearing parameters $\eta=0.05~{\rm eV}$, $\eta=0.1~{\rm eV}$, and $\eta=0.5~{\rm eV}$ in Fig.~\ref{fig:ilust_h}.
At this momentum transfer value, the DSF is primarily determined by collective plasmon scattering.
Panel (a) clearly illustrates the ambiguity and difficulty of assessing the $\eta$-convergence of the DSF at a given set of simulation parameters (cutoff energy, number of bands, $k$-point grid, etc.). 
Increasing $\eta$ from $0.05~{\rm eV}$ to $0.1~{\rm eV}$ significantly reduces the magnitude of fluctuations in the DSF. The results computed at $\eta=0.05~{\rm eV}$ effectively encompass those generated at $\eta=0.1~{\rm eV}$. When $\eta$ is further increased to $0.5~{\rm eV}$, we obtain a much smoother curve. This curve remains within the range covered by the data at $\eta=0.05~{\rm eV}$, but it shows reduced values around the plasmon region and in the low-frequency range ($\omega \lesssim 15~{\rm eV}$) to the degree that it barely overlaps with the results at $\eta=0.1~{\rm eV}$. In contrast, as we show in Fig.~\ref{fig:ilust_h}(b), the ITCFs computed using $\eta=0.05~{\rm eV}$ and $\eta=0.1~{\rm eV}$ are in excellent agreement with each other, while the ITCF calculated using  $\eta=0.5~{\rm eV}$ significantly deviates from both of them.
This exemplifies the suitability of the ITCF for the $\eta$-convergence test as the smoother DSF at $\eta=0.5~{\rm eV}$ comes at the cost of a significant unphysical bias that can be clearly detected in $F(\mathbf{q},\tau)$.

For the $\eta$-convergence test in the imaginary time domain, it is more appropriate to use the shifted ITCF values:  
\begin{equation}
    \widetilde F(\vec q, \tau)=F(\vec q, \tau)-F(\vec q, \tau=0).
    \label{eq:tilde_ITCF}
\end{equation}
This adjustment is necessary because TDDFT data for a fixed configuration of ions do not capture the quasi-elastic part of the DSF \cite{Mo_prb_2020, Gregori_PRE_2003} (e.g., caused by phonons or ionic plasma oscillations), which translates to a constant shift of the ITCF that does not depend on $\tau$~\cite{Bellenbaum_prr_2025}. Also, TDDFT does not allow for the adequate computation of the $\omega\to0$ limit of the DSF due to finite size effects \cite{moldabekov2025abinitiodensityfunctional}. A similar effect is well known, e.g., for the dynamic Onsager coefficients from the Kubo-Greenwood method \cite{Melton_POP_2024}. The impact of these finite-size effects at the $\omega\to0$ limit of the DSF is also manifested as an uncertainty in a constant shift in $F(\vec q, \tau)$ \cite{moldabekov2025abinitiodensityfunctional}.

\begin{figure}[t!]
  \centering
  \begin{tabular}{ c }
    \includegraphics[width=.45\linewidth]{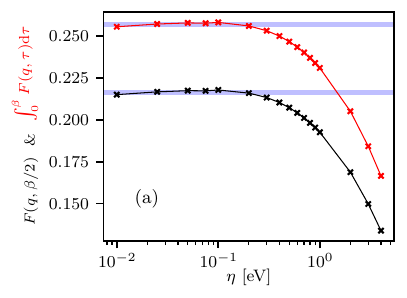}
  \end{tabular}%
    \begin{tabular}{ c }
    \includegraphics[width=.45\linewidth]{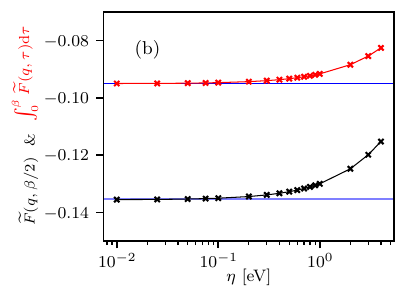}
 \end{tabular}%
  \caption{ Panel (a): dependence of the ITCF minimum, $F(\mathbf{q}, \tau=\beta/2)$, and the area under the ITCF, $\int_0^{\beta} F(\mathbf{q}, \tau) \, {\mathrm{d}}\tau$, on the Lorentzian smearing parameter $\eta$. Panel (b):  dependence of the minimum of the shifted ITCF (defined in Eq.(\ref{eq:tilde_ITCF})), $\widetilde F(\mathbf{q}, \tau=\beta/2)$, and the area under the shifted  ITCF, $\int_0^{\beta} \widetilde F(\mathbf{q}, \tau) \, {\mathrm{d}}\tau$, on the Lorentzian smearing parameter $\eta$. The results are presented for solid density hydrogen with $T=4.8 ~{\rm eV}$ and $\rho=0.08 ~{\rm g/cc}$ at $q=0.946~\textup{~\AA}^{-1}$.} \label{fig:ilust_h_itcf}
\end{figure}

We present a more detailed and systematic $\eta$-convergence test of the ITCF for solid density hydrogen at a temperature of $T=4.8~{\rm eV}$ and a momentum transfer of $q=0.946~\textup{~\AA}^{-1}$ in Fig. \ref{fig:ilust_h_itcf}. In panel (a), we show the dependence of the ITCF minimum, $F(\mathbf{q}, \tau=\beta/2)$, on $\eta$. As stated above, this minimum directly follows from the detailed balance relation in thermal equilibrium. Additionally, we display the $\eta$-dependence of the area under the ITCF, which corresponds to the static electronic density response scaled by a constant factor \cite{Dornheim_mre_2023},   $\chi(\vec q)=-n_e\int_0^{\beta} F(\mathbf{q}, \tau) \, {\mathrm{d}}\tau$. We observe that the minimum of the ITCF and the area under the ITCF both exhibit a weak dependence on $\eta$ for $\eta \leq 0.1~{\rm eV}$, while there is a strong dependence for $\eta > 0.1~{\rm eV}$. Due to the previously mentioned finite-size effects in the DSF at small frequencies, there is a small reduction in both the ITCF minimum and the area under the ITCF as $\eta$ decreases below $0.05~{\rm eV}$. Therefore, we consider $\widetilde F(\mathbf{q}, \tau=\beta/2)$ and $\int_0^{\beta} \widetilde F(\mathbf{q}, \tau) \, {\mathrm{d}}\tau$ in Fig. \ref{fig:ilust_h_itcf}(b), as they are not afflicted by these artifacts and clearly exhibit convergence with respect to $\eta$ at $\eta\leq 0.1~{\rm eV}$. 
From this, we draw the important intermediate conclusion that TDDFT calculations for the ITCF can achieve convergence with respect to the parameter $\eta$. Therefore, the largest $\eta$  within the convergence range provides the DSF that exhibits the smallest magnitude of numerical fluctuations for a given set of simulation parameters without showing any detectable bias in the Laplace transform, and the corresponding DSF data is referred to as \textit{the reference DSF}.

We emphasize that the generation of the DSFs with various $\eta$ takes place during the post-processing phase of the TDDFT simulation\cite{TIMROV2015460, Baczewski_PRL_2016, White_Electronic_Structure_2025} and requires negligible computational time. Therefore, the search for the reference DSF using the $\eta$-convergence test in the imaginary time domain adds negligible computational overhead.
To further reduce the level of non-physical fluctuations in the DSF, there now are two options: (i) one might increase the number of $k$-points and/or the system size, which are both computationally demanding and, indeed, can even be prohibitive; here, we advocate for a second alternative strategy (ii) in using a constraints-based filter that removes the artificial ringing from the DSF without the need for additional TDDFT efforts, and without the introduction of any significant bias as it occurs in the $\eta$-smearing for larger values of $\eta$.

\section{Constraints-based attenuation of narrow-band fluctuations}\label{s:constr}

The convergence of TDDFT results with respect to $\eta$ in the imaginary time domain indicates that one can write the TDDFT data for the DSF as a sum of a physical part of the DSF, $S_p(\vec{q},\omega)$, and the quasi-periodic non-physical fluctuations around it, $S_n(\vec{q},\omega)$, i.e.,
\begin{equation}
    S_{\rm TDDFT}(\vec{q},\omega)=S_p(\vec{q},\omega)+S_n(\vec{q},\omega),
\end{equation}
with the corresponding ITCF decomposition:
\begin{equation}
    F_{\rm TDDFT}(\mathbf{q},\tau)=F_p(\mathbf{q},\tau)+F_n(\mathbf{q},\tau).
\end{equation}

\begin{figure*}[t!]
	\centering
    \begin{subfigure}{}
		\includegraphics[width=0.45\linewidth]{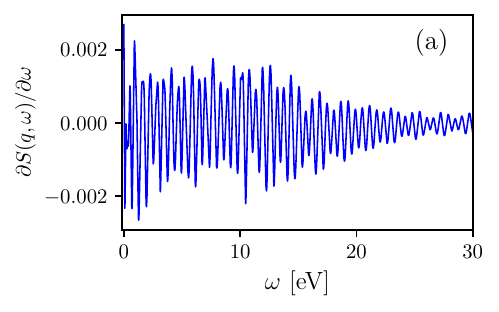}
	\end{subfigure}
     \begin{subfigure}{}
		\includegraphics[width=0.43\linewidth]{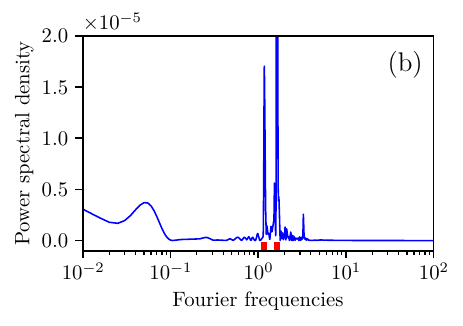}
	\end{subfigure}

\hspace{2.4em}%
     \begin{subfigure}{}
		\includegraphics[width=0.4\linewidth]{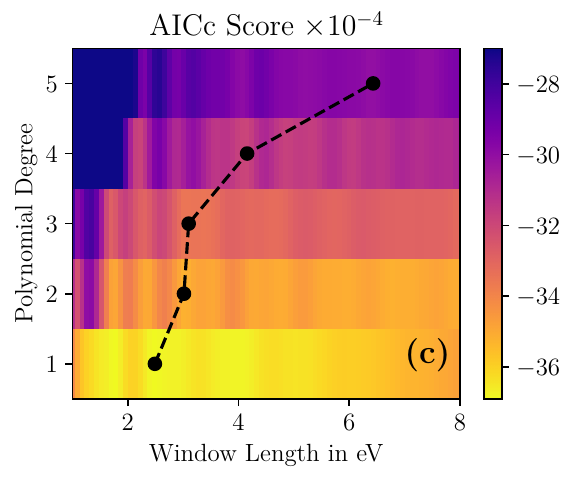}
	\end{subfigure}
     \begin{subfigure}{}
		\includegraphics[width=0.42\linewidth]{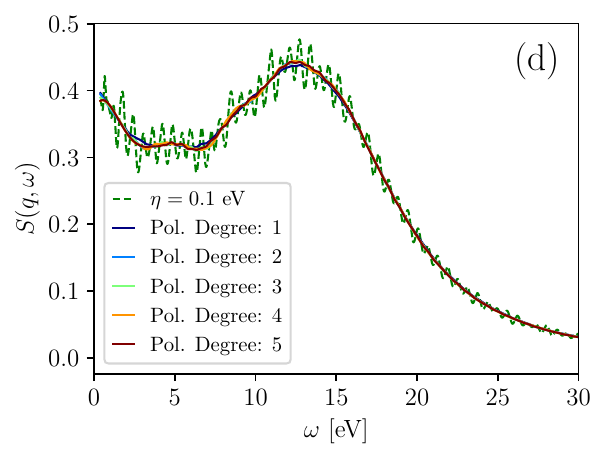}
	\end{subfigure}
\caption{(a) Derivative of the reference DSF, $\partial S(q,\omega)/\partial \omega$, with $\eta=0.1~{\rm eV}$ and at the same parameters as shown in Fig.~\ref{fig:ilust_h}. Panel (b) shows the power spectral density of $\partial S(q,\omega)/\partial \omega$ with the Fourier frequencies of the dominant narrowband fluctuations marked on the x-axis. Panel (c) displays the grid search over multiple window-size-polynomial-degree-combinations (color map) and the optimum window size for each polynomial degree (black symbols). In panel (d)  we compare the reference DSF with $\eta=0.1~{\rm eV}$ with the data obtained using the SG filer with different polynomial degrees and corresponding to them optimal window sizes as depicted in panel (a).}
	\label{fig:der_PSD}
\end{figure*}

The frequency of the quasi-periodic fluctuations in the DSF data can be easily identified by examining the spectrum of the derivative of DSF, $\partial S(q,\omega)/\partial \omega$. This is illustrated in the top panels of Fig. \ref{fig:der_PSD}. Panel (a) displays $\partial S(q,\omega)/\partial \omega$, while panel (b) shows the corresponding power spectrum. From Fig. \ref{fig:der_PSD}(b), we observe that the spectrum of fluctuations contains prominent narrowband components, with two major bands highlighted using red marks along the frequency axis.  Therefore, the task is to filter narrow-band fluctuations from the reference DSF data while ensuring a high degree of agreement with the reference ITCF. In general, when identifying the frequencies of narrowband fluctuations, complications that might be introduced by sharp DSF peaks associated with bound-state excitations can be circumvented by restricting the analysis of $\partial S(\vec q, \omega)/\partial \omega$ to a relatively featureless region of the reference spectrum and
use the smoothness of the derivative to impose a minimal stationary point condition for the filtering technique.


Filtering of non-physical fluctuations is a common procedure in data processing in fields such as analytical chemistry, geosciences, and medicine (see, for example, Refs. \cite{ACHARYA2016677, Schmid_ACS_2022}). 
Among the large variety of filters readily available for application, we use the Savitzky-Golay (SG) filter~\cite{SG_1964}, which is commonly used in spectroscopy. The parameters of the SG filter are the polynomial order, which defines the degree of the polynomial fit, and the window length, which specifies the number of data points in the moving window. 
The choice of the optimal window length that attenuates narrow-band fluctuations at a given polynomial degree can be performed automatically using various model selection metrics (scoring algorithms).
We use the corrected Akaike Information Criterion (AICc) \cite{AICC} to select filter parameters, which balances how well the filter fits the data with a penalty for unnecessary complexity. 
To address the specific concerns related to our application, we have introduced two additional penalties. The first is a frequency-domain penalty designed to prevent solutions from retaining contributions around specified target frequencies, which correspond to the dominant components of narrow-band fluctuations. The second is a second-derivative penalty that increases with the local curvature of the filtered signal. 
The application of this metric is illustrated in panel (c) of Fig. \ref{fig:der_PSD}, which displays the grid search involving different combinations of window sizes and polynomial degrees, as well as the optimal window size for each polynomial degree. In panel (d), the DSFs obtained using the identified optimal SG filter parameters are compared to the reference DSF data.  The data shown in panel (d) reveal that the results for $S_p(\vec q, \omega)$ with attenuated narrow-band fluctuations closely match each other, demonstrating the effectiveness and reliability of the applied strategy. The used postprocessing filter for the considered case is publicly available \cite{NoiseFilterTDDFT}.

\begin{figure*}[t!]
	\centering
    \begin{subfigure}{}
    \includegraphics[width=0.26\linewidth]{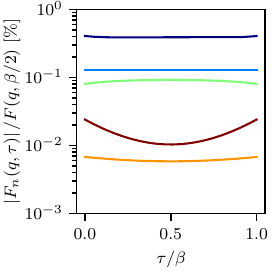}
	\end{subfigure}
    \begin{subfigure}{}
		\includegraphics[width=0.25\linewidth]{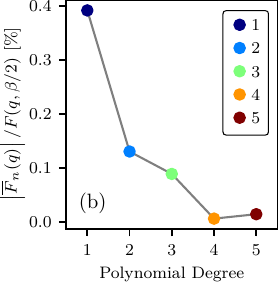}
	\end{subfigure}
    \begin{subfigure}{}
		\includegraphics[width=0.255\linewidth]{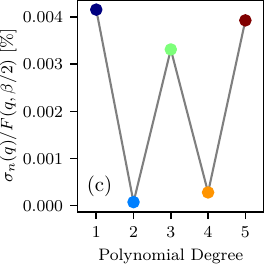}
	\end{subfigure}
    \caption{Absolute value of the residual ITCF $F_n(q,\tau)$ [(a)],  the magnitude of the mean $ \bar{F}_n(q)$ [(b)], and the RMS $ \sigma_n(q) $ [(c)] at different values of the polynomial degree and corresponding optimal window length. The results are for solid density hydrogen with $T=4.8 ~{\rm eV}$ and $\rho=0.08 ~{\rm g/cc}$ at $q=0.946~\textup{~\AA}^{-1}$.}
	\label{fig:sg_par}
\end{figure*}

To ensure a minimal bias, a solution for $S_p(\vec q,\omega)$ must satisfy additional conditions in the imaginary time domain.
As the first conditions, we consider the class of $S_p(\vec{q},\omega)$ solutions with:
\begin{eqnarray}\label{eq:Fn}
\frac{F_n(\mathbf{q},\tau)}{F_{\rm TDDFT}(\mathbf{q},\tau)} = \frac{\int_{-\infty}^\infty \textnormal{d}\omega\ S_n(\mathbf{q},\omega)\ e^{-\tau\omega}}{\int_{-\infty}^\infty \textnormal{d}\omega\ S_{\rm TDDFT}(\mathbf{q},\omega)\ e^{-\tau\omega}}=\epsilon(\tau)\ll 1,
\end{eqnarray}
which means that the magnitude of $F_n(\mathbf{q},\tau)$ is negligibly small compared to $F_{\rm TDDFT}(\mathbf{q},\tau)$ at all $\tau$ values and  corresponds to the requirement of the close agreement between the ITCFs of $S_{\rm TDDFT}(\vec{q},\omega)$ and $S_p(\vec{q},\omega)$:
\begin{eqnarray}\label{eq:Fs}
\int_{-\infty}^\infty \textnormal{d}\omega\
 S_{\rm TDDFT}(\mathbf{q},\omega)\ e^{-\tau\omega}\xlongequal{\epsilon(\tau)\to0}\int_{-\infty}^\infty \textnormal{d}\omega\  S_p(\mathbf{q},\omega)\ e^{-\tau\omega}.
\end{eqnarray}

The second condition stems from the requirement that $F_n(\mathbf{q}, \tau)$ contributes negligibly to the curvature of $ F_{\rm TDDFT}(\mathbf{q}, \tau)$. This can be expressed in multiple ways; here, we utilize the normalized root-mean-square deviation (RMS) of $F_n(\mathbf{q}, \tau)$ from its mean value $ \bar{F}_n(\mathbf{q})$ and compare it with the smallest value of the reference ITCF given by $F_{\rm TDDFT}(\vec q, \tau=\beta/2)$. The corresponding condition reads:
\begin{equation}\label{eq:RMS}
    \sigma_n(\vec q)=\frac{1}{\tau}\left(\int_0^{\tau} \left(F_n(\vec q, \tau)-\bar{F}_n(\mathbf{q})\right)^2\mathrm{d}\tau\right)^{1/2}\ll F_{\rm TDDFT}(\vec q, \tau=\beta/2).
\end{equation}
The requirement of minimizing the effect of filtering on the ITCF curvature, along with the condition outlined in Eq.~(\ref{eq:Fn}), ensures that the frequency moments of the DSF are impacted as little as possible, since these moments can be expressed as the derivatives of the ITCF with respect to the imaginary-time argument $\tau$ \cite{Dornheim_prb_2023,Dornheim_mre_2023,tolias2025exactseriesexpansionfrequency}.
It is important to note that a converged ITCF alone does not provide a unique determination of the DSF, since the inverse mapping (analytic continuation) from the ITCF in the imaginary-time domain to the DSF on the real-frequency axis is subject to practical numerical instabilities \cite{JARRELL1996133, PhysRevB.95.014102}.
Nevertheless, convergence of the ITCF is a necessary criterion for any reliable method for interpreting XRTS measurements at extreme conditions, as it automatically ensures convergence of a number of physical characteristics, including the frequency moments of the DSF~\cite{Dornheim_prb_2023}.

In Fig. \ref{fig:sg_par}, for the DSF results from Fig. \ref{fig:der_PSD} (d), we show the magnitude of the residual ITCF $F_n(q, \tau)$ in panel (a), the mean $\bar{F}_n(q)$ in panel (b), and the RMS $\sigma_n(q)$ in panel (c). Fig. \ref{fig:sg_par} demonstrates that all identified optimal combinations of window and polynomial degrees meet the conditions outlined in equations (\ref{eq:Fn}) and (\ref{eq:RMS}) with a high degree of accuracy. Specifically, we have $\bar F_n(q) / F_{\rm TDDFT}(\vec q, \beta/2) < 0.4\%$ and $\sigma_n(q) / F_{\rm TDDFT}(\vec q, \beta/2) \leq 0.0042\%$.

The close alignment of the obtained curves and their strict compliance with the specified constraints in the imaginary time domain indicate that the attenuation of narrow-band fluctuations produces a reliable representation of \(S_p(\vec{q}, \omega)\). In summary, regardless of the filtering strategy employed, the key criteria are (1) efficient attenuation of narrowband fluctuations and (2) minimal impact on the ITCF.

\begin{figure*}[t!]
	\centering
    \begin{subfigure}{}
		\includegraphics[width=0.9\linewidth]{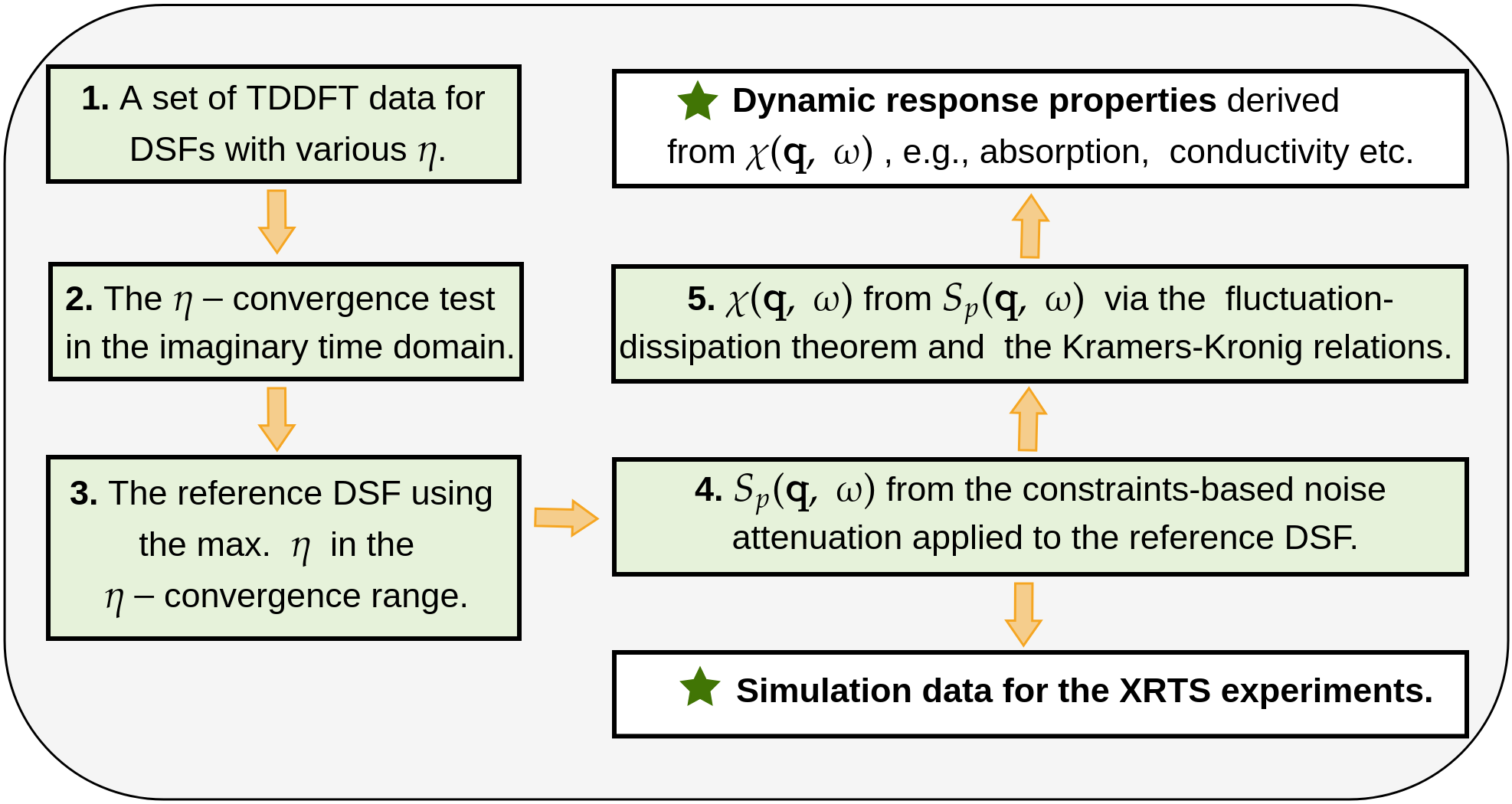}
	\end{subfigure}
\caption{Workflow for applying the $\eta$-convergence test in the imaginary time domain combined with the constraints-based attenuation of narrow-band fluctuations. For the first step, the generation of the DSFs with various $\eta$  is realized at the post-processing phase of the TDDFT simulation and requires negligible computational time.}
	\label{fig:workflow}
\end{figure*} 

We stress that the use of the constraints-based filtering method extends beyond modeling the DSF, as the latter is directly related to the imaginary component of the dynamic density response function via the fluctuation-dissipation theorem \cite{quantum_theory}: 
\begin{equation}\label{eq:fluc_dis}
    S(\vec q, \omega)=-\frac{\hbar^2}{n}~\frac{1}{1-e^{-\hbar \omega/k_BT}}~{\rm Im}\left[\chi(\vec q, \omega)\right],
\end{equation}
where $n$ is the electronic number density. 
Once $S_p(\vec q, \omega)$ is determined, the fluctuation-dissipation theorem thus directly provides the imaginary part of the corresponding dynamic density response function. Furthermore, the Kramers-Kronig relations enable the computation of its real part~\cite{quantum_theory} and corresponding complete knowledge of $\chi(\mathbf{q},\omega)$ gives one access to a host of dynamic dielectric properties, such as absorption coefficient and conductivity (see Sec.~\ref{ss:conection}).  The workflow for applying the $\eta$-convergence test in the imaginary time domain, in conjunction with the constraints-based filtering of narrow-band fluctuations, is summarized in Fig.~\ref{fig:workflow}.

Instead of starting from the DSF, one can also begin with any other dynamic dielectric property, e.g, conductivity. Then, by applying linear-response theory and the fluctuation-dissipation theorem, it is possible to derive the reference DSF data that are afflicted by the unphysical narrow-band fluctuations. Subsequently, the constraints-based filtering method can be used to determine an unbiased solution for $S_p(\vec{q}, \omega)$. Using $S_p(\vec{q}, \omega)$, one can then propagate backwards to compute smooth and physical curves for any relevant dynamic response properties.

\section{Examples of practical applications}\label{s:results}
\subsection{Solid density hydrogen}\label{s:hydrogen}

For solid-density hydrogen at $T = 4.8~\text{eV}$, $\rho = 0.08~\text{g/cc}$, and $q = 0.946~\text{\AA}^{-1}$, we adopt the average of the results obtained using SG filters with different polynomial degrees and their corresponding optimal window lengths as the solution for $S_p(\vec{q}, \omega)$. 
Further technical details of the TDDFT simulations are given in Sec.~\ref{s:dft_info}.

\begin{figure*}[t!]
	\centering
    \begin{subfigure}{}
		\includegraphics[width=0.4\linewidth]{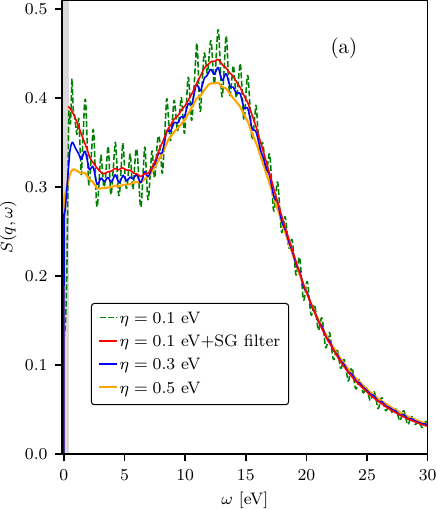}
		\label{fig:subfigA}
	\end{subfigure}\hspace{-0.8em}%
    \begin{subfigure}{}
		\includegraphics[width=0.465\linewidth]{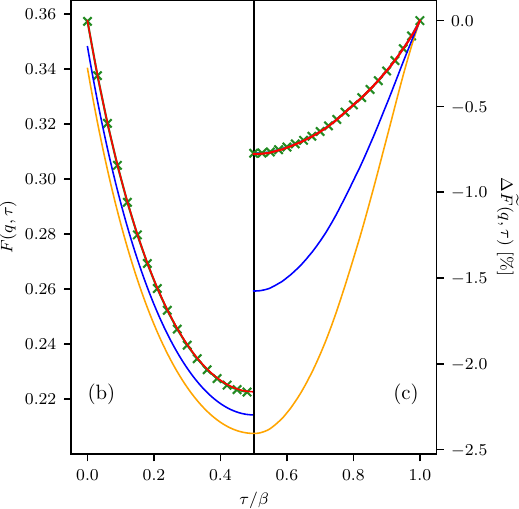}
	\end{subfigure}
    
    \hspace{-3.25em}%
    \begin{subfigure}{}
		\includegraphics[width=0.8\linewidth]{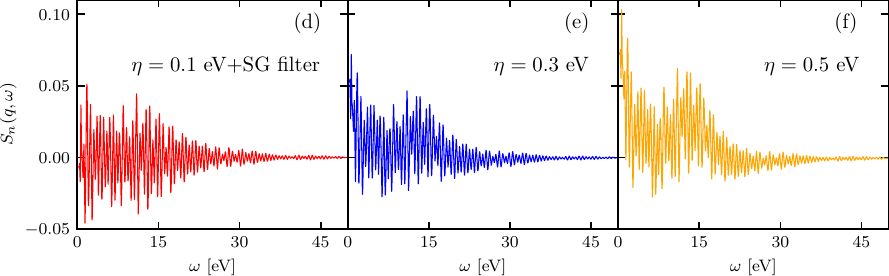}
	\end{subfigure}
\caption{(a) TDDFT results for the DSF of solid density hydrogen with $T=4.8 ~{\rm eV}$ and $\rho=0.08 ~{\rm g/cc}$ at $q=0.946~\textup{~\AA}^{-1}$. Panel (b) shows the corresponding ITCF results.  In panel (c), we quantify the accuracy of the TDDFT results compared to exact PIMC data~\cite{Dornheim_MRE_2024,Bellenbaum_prr_2025} using the deviation measure (\ref{eq:delta_F}). Shown is the difference between the DSF data computed using $\eta=0.1~{\rm eV}$ and the DSF results obtained using the constraints-based attenuation of narrow-band fluctuations and also using values of $\eta=0.3~{\rm eV}$ and $\eta=0.5~{\rm eV}$. }
	\label{fig:filt_h}
\end{figure*}

In Fig.~\ref{fig:filt_h}, we compare $S_p(\vec q, \omega)$ and the reference DSF computed with $\eta = 0.1~\text{eV}$ to the DSF data obtained with $\eta = 0.3~\text{eV}$ and $\eta = 0.5~\text{eV}$. From the top panels (a) and (b) of Fig.~\ref{fig:filt_h}, we see that, in contrast to $S_p(\mathbf{q},\omega)$, the smoothing of the DSF by increasing $\eta$ to $0.3~\text{eV}$ and $0.5~\text{eV}$—values outside the $\eta$-convergence range—leads to a significant suppression of the DSF features at $\omega \lesssim 15~\text{eV}$ and causes substantial discrepancies between the corresponding ITCF results and the reference $F(\vec{q}, \tau)$ data.


As an independent check of the TDDFT data, we show the deviation of the TDDFT results for the ITCF  from the exact path integral quantum Monte Carlo (PIMC) data ~\cite{Dornheim_MRE_2024} in Fig.~\ref{fig:filt_h}(c), where the relative deviation is estimated using:
\begin{equation}\label{eq:delta_F}
    \Delta \widetilde F(\vec q,\tau) [\%]=\frac{\widetilde F_{\rm TDDFT}(\vec q,\tau)-\widetilde F_{\rm PIMC}(\vec q,\tau)}{\widetilde F_{\rm PIMC}(\vec q,\tau=\beta/2)} \times 100,
\end{equation}
where $\widetilde  F_{\rm PIMC}(\vec q,\tau)$ is the shifted ITCF data from PIMC.  
The TDDFT data with $\eta=0.1~{\rm eV}$ and the corresponding filtered results are in very good agreement with the exact PIMC results with $\Delta \widetilde F(\vec q,\tau)<0.75~\%$. The growing deviation between the TDDFT and PIMC results as $\tau$ approaches $\beta/2$ is attributed to the adiabatic exchange correlation kernel approximation in our TDDFT simulations~\cite{moldabekov2025abinitiodensityfunctional}. 
As expected, the increase in $\eta$ to the values outside of the $\eta$-convergence range leads to an increase in the deviation of the TDDFT data from the exact PIMC results.

 \begin{figure*}[t!]
	\centering
    \begin{subfigure}{}
		\includegraphics[width=0.45\linewidth]{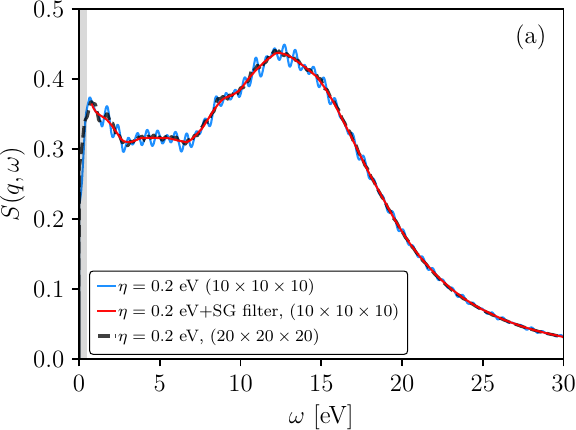}
	\end{subfigure}
     \begin{subfigure}{}
		\includegraphics[width=0.45\linewidth]{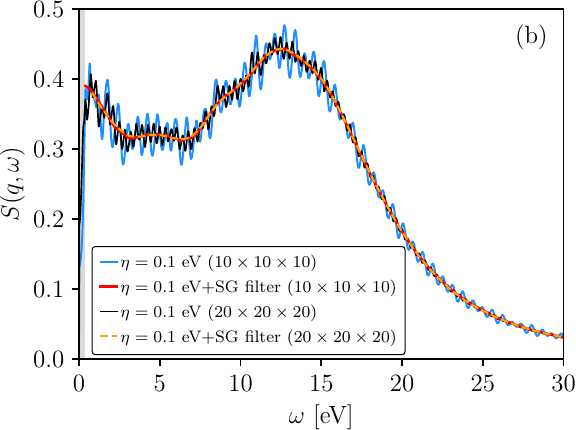}
	\end{subfigure}
\caption{Comparison of the DSF computed using $10\times10\times10$ and $20\times20\times20$ k-point grids for hydrogen at $T = 4.8~\mathrm{eV}$ and $\rho = 0.08~\mathrm{g/cc}$, at $q = 0.946~\textup{\AA}^{-1}$. The results are shown for $\eta = 0.2~\mathrm{eV}$ in panel~(a) and for $\eta = 0.1~\mathrm{eV}$ in panel~(b).
}\label{fig:r1}
\end{figure*}

In Fig.~\ref{fig:filt_h}(d), we present the difference between the DSF data before and after applying the constraints-based elimination of narrow-band
fluctuations, i.e., $S_n(\vec{q},\omega) = S(\vec{q},\omega) - S_p(\vec{q},\omega)$. 
For comparison, in in Fig.~\ref{fig:filt_h}(e) and in Fig.~\ref{fig:filt_h}(f), we also show the difference between the DSF data computed using $\eta=0.1~{\rm eV}$ and the DSF results smoothed by increasing $\eta$ to $0.3~{\rm eV}$ and $0.5~{\rm eV}$.
We observe that the constraints-based filtering produces $S_n(\vec{q}, \omega)$ fluctuating symmetrically around zero. In contrast, the commonly used method of smoothing by increasing $\eta$ leads to a loss of physically relevant information, and, hence, biased results for the DSF.

The discussed quasi-periodic fluctuations are a direct consequence of the finite k-point grid.
We demonstrate this by comparing data from calculations using $20\times20\times20$ and $10\times10\times10$ k-point grids for $\eta=0.1~{\rm eV}$ and $\eta=0.2~{\rm eV}$.
In Fig.~\ref{fig:r1} (a), one can see that, at $\eta=0.2~{\rm eV}$, the DSF computed using the $20\times20\times20$ k-point grid (black dashed line) is substantially smoother than the results obtained with the $10\times10\times10$ grid, indicating a significant reduction of finite-k discretization effects. When the constraints-based filtering of narrow-band fluctuations is applied to the data generated with the $10\times10\times10$ grid, the resulting DSF (red solid line) closely matches the result computed using a $20\times20\times20$ k-point grid.

From panel~(a) of Fig.~\ref{fig:r1}, we observe that the reduction of $\eta$ from $\eta = 0.2~\mathrm{eV}$ to $\eta = 0.1~\mathrm{eV}$ leads to the appearance of quasi-periodic fluctuations even for the $20\times20\times20$ k-point grid, although with significantly smaller magnitude than in the $10\times10\times10$ case. This means that decreasing $\eta$ inevitably amplifies finite-k discretization effects for any finite k-point grid. Importantly, after applying the constraints-based procedure, the DSFs obtained from the two k-point grids (red solid line for $10\times10\times10$, orange dashed line for $20\times20\times20$) are smooth and in very close agreement with each other. This agreement indicates that, at $\eta = 0.1~\mathrm{eV}$, the results obtained using the proposed approach are effectively converged with respect to k-point density.

\begin{figure*}[t!]
	\centering
    \begin{subfigure}{}
		\includegraphics[width=0.44\linewidth]{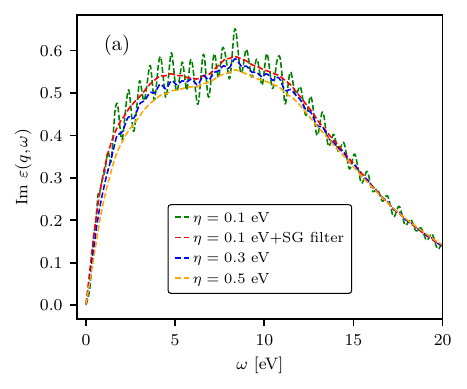}
	\end{subfigure}
     \begin{subfigure}{}
		\includegraphics[width=0.447\linewidth]{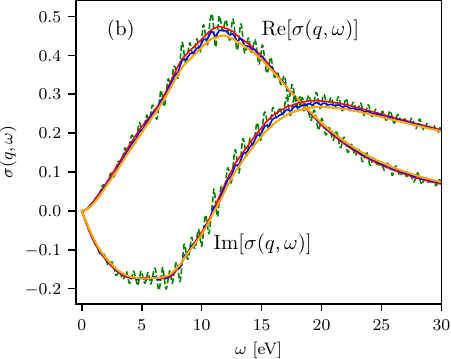}
	\end{subfigure}
\caption{(a) Imaginary part of the dynamic dielectric function and (b) dynamic conductivity of solid density hydrogen with $T=4.8 ~{\rm eV}$ and $\rho=0.08 ~{\rm g/cc}$ at $q=0.946~\textup{~\AA}^{-1}$. 
}\label{fig:absorbtion}
\end{figure*} 

As discussed in Sec.\ref{s:constr}, the fluctuation-dissipation theorem and the Kramers-Kronig relations link the DSF with other dynamic response properties (see Sec.~{\ref{ss:conection}}). To illustrate the application of the $\eta$-convergence test in the imaginary time domain in conjunction with the constraints-based attenuation of narrow-band fluctuations, we present results for the Imaginary part of the dynamic dielectric function (which, e.g., describes electromagnetic wave absorption \cite{alexandrov1984principles}) and dynamic conductivity of dense hydrogen in Fig.~\ref{fig:absorbtion}. We present the results based on $S_p(\vec{q},\omega)$, which were obtained through constraints-based filtering, along with the data generated from the reference DSF with $\eta=0.1~{\rm eV}$. Additionally, we compare to the results computed using $\eta=0.3~{\rm eV}$ and $\eta=0.5~{\rm eV}$.
In Fig. \ref{fig:absorbtion}, we see that the most notable discrepancies between the data based on $S_p(\vec{q},\omega)$ and the results obtained using $\eta=0.3~{\rm eV}$ and $\eta=0.5~{\rm eV}$ occur near their respective maximum values.

\begin{figure*}[t!]
	\centering
    \begin{subfigure}{}
		\includegraphics[width=0.9\linewidth]{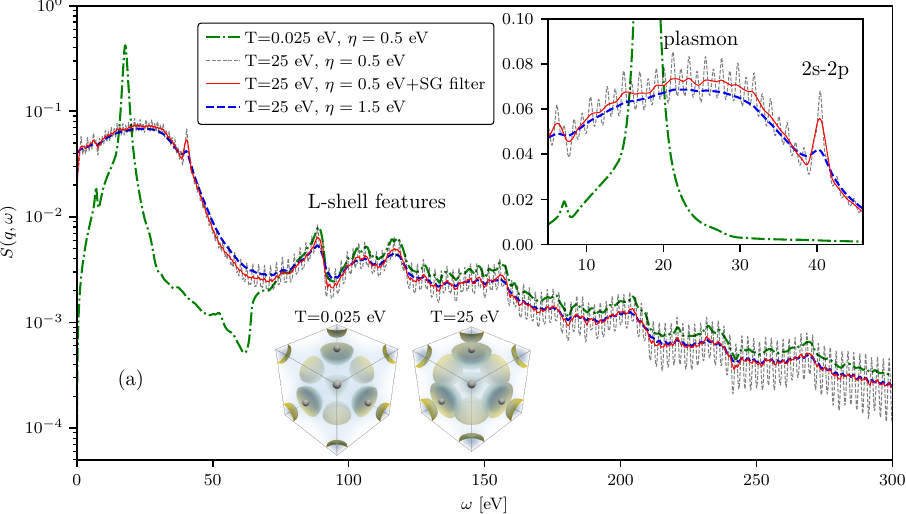}
	\end{subfigure}

 \hspace{-1em}%
  \begin{tabular}{ c }
    \includegraphics[width=.45\linewidth]{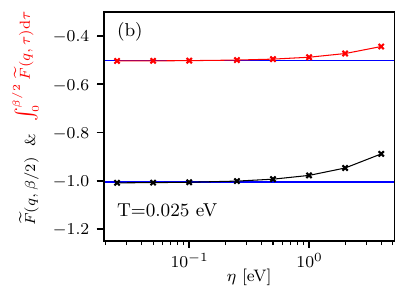}
  \end{tabular}%
    \begin{tabular}{ c }
    \includegraphics[width=.45\linewidth]{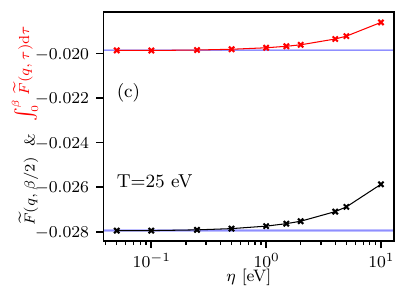}
 \end{tabular}%
\caption{(a) TDDFT results for the DSF of Al with $T=0.025~{\rm eV}$ and for isochorically heated electrons with $T=25~{\rm eV}$. The electronic isosurfaces in fcc aluminum (boxes at the bottom of panel (a)) demonstrate the spread of electron density into the interstitial region as a result of isochoric heating~\cite{Moldabekov_jctc_2024}.  
The bottom panels (b) and (c) display the $\eta$-convergence tests in the imaginary time domain using the minimum of the shifted ITCF at $\tau=\beta/2$ and the area under the shifted ITCF.   
The presented data are for the wavenumber $q= 1.0~\textup{~\AA}^{-1}$ directed along the [100] direction relative to the crystal structure.}
	\label{fig:Al_low}
\end{figure*}

\begin{figure*}[t!]
	\centering
    \begin{subfigure}{}
		\includegraphics[width=0.9\linewidth]{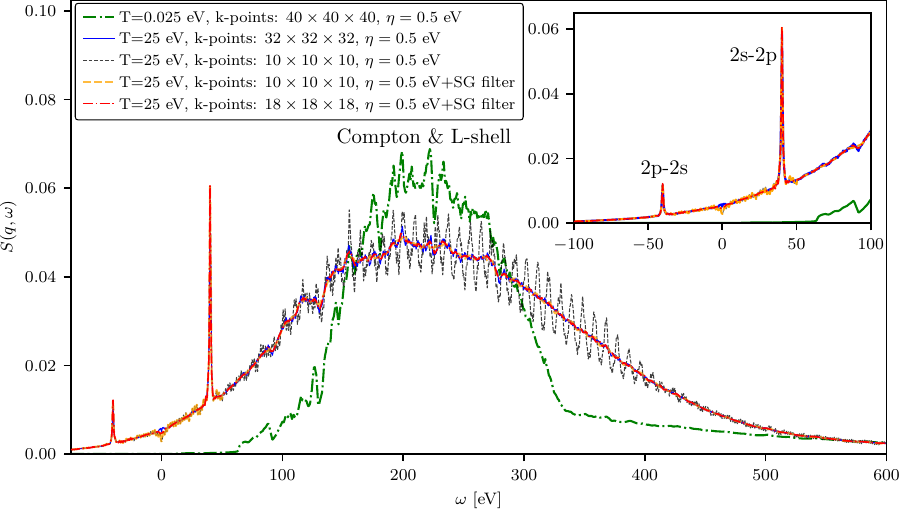}
	\end{subfigure}
\caption{TDDFT results for the DSF of Al with $T=0.025~{\rm eV}$ and isochorically heated electrons with $T=25~{\rm eV}$ at the wavenumber $q= 7.75~\textup{~\AA}^{-1}$ directed along the [100] direction relative to the crystal structure.}
	\label{fig:al_high}
\end{figure*}

\subsection{Aluminum with a face-centered cubic crystal structure}\label{s:hydrogen}

As a second example of applying the $\eta$-convergence test in the imaginary time domain, along with the constraints-based filtering of narrow-band fluctuations, we examine Al with a face-centered cubic (fcc) crystal structure. This material is frequently used as a testbed for new diagnostic techniques of electronic structure properties \cite{Preston_APL_2019, Gawne_PRB_2024} and for developing computational methods \cite{Panholzer_PRL_2018, Kononov2023, Witte_PRL, gawne2025orientationaleffectslowpair}.  

In Fig.~\ref{fig:Al_low}(a), we present TDDFT results for the DSF of electrons in fcc aluminum with 11 valence electrons per atom at momentum transfer $q= 1.0~\textup{~\AA}^{-1}$ along the [100] direction relative to the crystal structure. We analyze two cases: ambient conditions at a temperature of  $T=0.025~\rm ~{\rm eV}$ ($T\sim290~{\rm K}$) and an isochorically heated state at $T=25~\rm ~{\rm eV}$.  The results of the analysis on the $\eta$-convergence in the imaginary time domain are shown in Figs.~\ref{fig:Al_low}(b) and \ref{fig:Al_low}(c) for $T=0.025~\rm ~{\rm eV}$ and for $T=25~\rm ~{\rm eV}$, respectively, showing both the dependence on $\eta$ of the minimum of the shifted ITCF and the area under the shifted ITCF. We find that $\eta$-convergence is reached at $\eta\leq 0.5~{\rm eV}$. 
The DSF results computed with $\eta=0.5~{\rm eV}$ thus serve as the reference data for the application of the constraints-based SG filtering at $T=25~{\rm eV}$. In the isochorically heated scenario at a temperature of $T=25~\rm eV$, there is a significant computational bottleneck due to a large number of partially occupied orbitals and the requirement of considerable cutoff energy to describe thermal excitation from 2s and 2p states (L-shell) correctly. Considering the cubic crystal structure with a fcc crystal lattice parameter of $ a = 4.05 \, \textup{~\AA}^{-1}$,  we used 100 orbitals per atom and $200~{\rm Ry}$ cutoff energy to capture enough partially occupied KS orbitals and to describe the effect of thermal excitations; with the smallest occupation number being less than $10^{-5}$ and the  $k$-point grid set to $10\times10\times10$ (see Sec.~\ref{s:filt_par} for the SG filter parameters).  Here, we show the cold data to demonstrate the changes in the DSF due to thermal excitations.  In the cold case, there are only six occupied KS orbitals per atom. With no computational issues, a DSF curve free of non-physical fluctuations at  $T=0.025~\rm ~{\rm eV}$  was generated using a $40 \times 40 \times 40$ $k$-point grid. 

In Fig.~\ref{fig:Al_low}(a), one can see a significant broadening and shift of the plasmon to larger energies in the DSF at $T=25~\rm eV$ compared to the cold case due to strong thermal excitations. 
The thermal excitations also generate a prominent 2s-2p transition feature, as discussed in Ref. \cite{baczewski2021predictionsboundboundtransitionsignatures}. In addition, we clearly see the L-shell features in both cold and hot cases. The reference DSF results at $T=25~\rm eV$ demonstrate significant fluctuations in both the plasmon and L-shell features. Increasing $\eta$ to $1.5~\rm eV$ reduces these fluctuations, but it also leads to a strong damping of the bound-bound 2s-2p feature. In contrast, the constraints-based SG filtering effectively attenuates narrow-band fluctuations across the entire considered frequency range while preserving the 2s-2p feature. 

Finally, to further show an enhancement in the efficiency of the TDDFT modeling of the DSF due to the presented workflow, we show the DSF for Al at $q= 7.75~\textup{~\AA}^{-1}$ at cold ($T=0.025~\rm eV$) and isochorically heated ($T=25~\rm eV$) conditions in Fig.~\ref{fig:al_high}. A momentum transfer of $q= 7.75~\textup{~\AA}^{-1}$ corresponds to a back-scattering geometry, where the detector is positioned at an angle greater than $ 90^\circ$ relative to the wavevector of the probing XFEL beam. This setup, which measures the so-called Compton feature, enables XRTS diagnostics to be performed without interfering with other diagnostic tools positioned in the forward direction, such as XRD and radiography. 

For $T=25~\rm eV$, we compare the data computed using $10 \times 10 \times 10$ and $32 \times 32 \times 32$ $k$-point grids (both with $\eta=0.5~{\rm eV}$). In addition, we show the results obtained by applying the constraints-based SG filtering to the data computed using $10 \times 10 \times 10$ $k$-point grid and $18 \times 18 \times 18$ $k$-point grid with $\eta=0.5~{\rm eV}$.
At $T=25~\rm eV$, we observe that the constraints-based SG filtering using $10 \times 10 \times 10$ $k$-point grid allows us to obtain DSF data with the  quality comparable to the simulations applying an $32 \times 32 \times 32$ $k$-point grid, where one can see a sharp and prominent 2s-2p feature and a broad Compton signal overlapping with the L-shell features. In this case, this constitutes more than a fifty-fold speedup, reducing the compute time from about $\sim 8.8\times10^5$  to $\sim 1.6\times10^4$ CPU hours~\footnote{This corresponds to worse than the ideal cubic scaling with respect to the k-point grid density, due to additional overhead associated with parallelization over 192 nodes with 96 CPUs per node.}. In practice, to describe the experimental XRTS measurement for a specific scattering angle, it is necessary to perform calculations for several wavenumbers surrounding the central value in order to account for the finite size of the detector \cite{Gawne_PRB_2024, Gawne_2025}. Additionally, the modeling of materials that are laser-heated or shock-compressed requires a series of simulations to consider variations in temperature and density within the target~\cite{doi:10.1063/1.4893146}, and also potentially an even wider range of densities and temperatures for a statistical analysis of the most probable set of conditions~\cite{Kasim_POP_2019}. Consequently, the present combination of the $\eta$-convergence test in the imaginary time domain along with constraints-based attenuation of narrow-band fluctuations has the potential for significant savings in computational time, and thereby to facilitate a more rigorous analysis of experimental measurements.

\section{Discussion}\label{sec:outlook}

State-of-the-art XFEL facilities and modern backlighter set-ups enable the investigation of material properties under new, hitherto unexplored conditions \cite{Kraus_Nature_2025, Descamps_sciadv, White_Nature_2025, Tilo_Nature_2023}. While advancements in experimental capabilities have greatly enhanced our understanding of high-pressure physics and laser-matter interaction, they also present new challenges for \textit{ab initio} simulation techniques. It has become crucial to improve the efficiency of \textit{ab initio} modeling due to the significant increase in computing power that is required to accurately describe the properties of materials under extreme conditions \cite{Dornheim_NatureCom_2025, Kononov2023, Ding_prl_2018, White_prl_2020}.  One key property measured directly in experiments is the XRTS spectrum. Modeling these measurements is a computationally highly intensive task due to strong thermal excitations, the peculiarities of the detectors used to capture the XRTS spectrum, and the variations in temperature and density within the experimental targets. Consequently, developing new computational strategies to improve the efficiency of XRTS modeling has become a crucial task for providing sustainable theoretical support for XRTS experiments, as well as for extending the benchmark and refinement of simplified models commonly used in areas such as radiation hydrodynamics. 

In this work, we developed an approach based on the one-to-one mapping between the electronic DSF $S(\mathbf{q},\omega)$ and the ITCF $F(\mathbf{q},\tau)$. This method combines the $\eta$-convergence test in the imaginary time domain with a constraints-based attenuation of narrow-band
fluctuations. This combined approach constitutes a highly efficient strategy for XRTS modeling, providing a speed-up of up to an order of magnitude and thus allowing for a significant reduction in the computational cost associated with the analysis of XRTS measurements for targets under extreme conditions. 
While convergence of the ITCF alone is not sufficient to uniquely determine the DSF, it does ensure the convergence of a number of physical characteristics \cite{Dornheim_mre_2023, TC_jcp_2025}, including the frequency moments of the DSF \cite{Dornheim_prb_2023, tolias2025exactseriesexpansionfrequency}, thereby providing a physically grounded basis for the computational strategy presented in this work.

For any given k-point grid, there exists a maximum attainable spectral resolution due to the limited amount of information about dynamic response properties contained in any finite k-point sampling, in contrast to the true thermodynamic limit with an infinite number of states. Besides this limitation, finite k-point sampling introduces narrow-band fluctuations. These fluctuations do not represent physical information but rather constitute an artifact whose amplitude increases as the broadening parameter is reduced. The presented approach improves the fidelity of the DSF by attenuating these identifiable finite-k-point-grid artifacts in a controlled manner, with a significantly smaller impact on resolved physical spectral features than increasing the broadening parameter, while preserving physical characteristics that follow from the ITCF. These benefits are achieved while substantially reducing the required computational cost.

The application of the presented workflow was demonstrated on the example of solid-density hydrogen with partially degenerate electrons under conditions relevant for IFE and laser-driven hydrogen jet experiments. A second example is given by isochorically heated aluminum with hot electrons within an fcc lattice structure of ions. This scenario is pertinent to the physics of transient states studied in experiments that utilize X-ray-driven heating \cite{Descamps_sciadv, Descamps2020, Matthieu2011, White2014, Sperling_PRL_2015} (also see discussions in Ref.~\cite{Moldabekov_prr_2024}).

By significantly reducing the computational burden associated with modeling XRTS experiments, the presented approach is expected to play a vital role in computational efforts aimed at supporting new experimental initiatives. In addition to XRTS modeling and interpretation, the presented workflow also enhances the efficiency of modeling a plethora of dynamic response properties that are related to the DSF through the fluctuation-dissipation theorem and the Kramers-Kronig relations. Instead of the DSF, the workflows can be initiated using any dynamic response property that can be expressed in terms of the dynamic density response function ~\cite{Schoerner_PRE_2023, Ramakrishna_2020, Hamann_PRB_2020}, such as dynamic conductivity from the Kubo-Greenwood method \cite{Schoerner_PRE_2023, Melton_POP_2024, CALDERIN2017118, PhysRevE.109.065304} (as outlined in Sec.~\ref{ss:KG_DSF}) or TDDFT \cite{ullrich2012time, Alex_prl_2025}.

\appendix
\renewcommand{\theequation}{A-\arabic{equation}}
\setcounter{equation}{0}
\renewcommand\thefigure{\thesection.\arabic{figure}}    
\setcounter{figure}{0}

\section{Appendix: Methods}\label{sec:methods}

\subsection{TDDFT simulation details}\label{s:dft_info}

The TDDFT simulations were carried out using Quantum ESPRESSO \cite{Giannozzi_2009, Giannozzi_2017, Giannozzi_jcp_2020, Carnimeo_JCTC_2023, TIMROV2015460}, using the Liouville-Lanczos (LL) method to compute the DSF.  A static (adiabatic) GGA approximation for the exchange-correlation (XC) kernel was employed in all calculations. For hydrogen, we utilized the pseudopotential $\textsc{h\_pbe\_v1.4.uspp.F.UPF}$ \cite{Garrity_CMS_2014} in conjunction with the PBE XC functional \cite{PhysRevLett.77.3865}. We considered a density value of $\rho = 0.08 ~\text{g/cc}$ and a temperature of $T = 4.8~\text{eV}$, corresponding to the Fermi temperature of the electrons~\cite{quantum_theory}. The size of the simulation box, which contains 14 protons, is set to $L = 6.641174 \textup{~\AA}$. This temperature leads to a partially degenerate state, with partial ionization~\cite{Bohme_PRL_2022,Bellenbaum_prr_2025,Hamann_PRR_2023,Bonitz_pop_2024}. The DSF results were averaged over 20 snapshots, using a Monkhorst-Pack $k$-point grid set to $10\times10\times10$. The convergence of the results with respect to the $k$-point grid was verified by testing an individual snapshot with different $k$-point grids: $8\times8\times8$, $10\times10\times10$, $14\times14\times14$, and $18\times18\times18$. The snapshots of protons were randomly selected from the particle trajectories generated by the KSDFT-based molecular dynamics simulations. 
To obtain $10^5$ Lanczos coefficients, we employed a bi-constant extrapolation scheme following the computed number of Lanczos iterations up to $N_{\rm iter}=2\times 10^4$ \cite{TIMROV2015460}. The calculations were executed with a cutoff energy of $75~\text{Ry}$ and $280$ KS orbitals.

For Al, we computed the DSFs in the [100] crystallographic direction and employed the GGA-level PBE functional along with the pseudopotential \textsc{Al.pbe-sp-van.UPF} \cite{PhysRevB.77.172102, PP_ESPRESSO_2022}, which accounts for 11 valence electrons, and an energy cutoff of 200 Ry. 
We employed a cubic simulation box containing the conventional fcc unit cell (4 atoms) with an fcc lattice parameter of $ a = 4.05 \, \textup{~\AA}^{-1} $~\cite{wyckoff1948crystal}.  
At the temperature of $ T = 0.025 \, \text{eV}$, we used a $40 \times 40 \times 40$ k-point grid. At the higher temperature of $T = 25 \, \text{eV}$, we employed two different k-point grids: $ 10 \times 10 \times 10 $ and $ 18 \times 18 \times 18 $. Additionally, at $ T = 25 \, \text{eV} $, we utilized 400 KS bands for the cell with 4 Al atoms. 

\subsection{Connection between different dynamic response functions}\label{ss:conection}

Here, we briefly outline the relationships between the DSF and the dynamic conductivity as well as the dynamic dielectric function, which are used to compute the results presented in Section \ref{s:hydrogen}. For more details, see, e.g., Refs.\cite{ginzburg1970propagation, alexandrov1984principles}. The DSF is connected to the imaginary part of the density response function through the fluctuation-dissipation theorem, as indicated in Eq. (\ref{eq:fluc_dis}). This means that the fluctuation-dissipation theorem and the Kramers-Kronig
relations allow one to compute the density response function $\chi(\vec q,\omega)$ from the DSF. The dynamic dielectric function can be expressed in terms of $\chi(\vec q,\omega)$ as
\begin{eqnarray}\label{eq:di_f}
    \varepsilon^{-1}(\vec q, \omega)=1+\frac{4\pi}{q^2}\chi(\vec q, \omega),
\end{eqnarray}
which also allows one to compute the dynamic conductivity using the relation
\begin{eqnarray}\label{eq:di_f2}
    \varepsilon(\vec q, \omega)=1+\frac{4\pi i}{\omega}\sigma(\vec q, \omega).
\end{eqnarray}


    

\subsection{The $\eta$-convergence test using the Kubo-Greenwood dynamic conductivity}\label{ss:KG_DSF}

The Kubo-Greenwood method is commonly employed to characterize the dynamic transport properties of materials under extreme conditions in the optical limit \cite{Hu_pop_2025, Melton_POP_2024, Maximilian_2023, CALDERIN2017118, PhysRevE.109.065304}. Given the importance of this approach for such applications as warm dense matter and dense plasmas, we outline the steps to implement the $\eta$-convergence test in the imaginary time domain alongside a constraints-based filtering starting from the Kubo-Greenwood conductivity $\sigma_{\rm KG}(\omega)$. In this context, we consider the averaged value of $\sigma_{\rm KG}(\omega)$ defined as the trace of the conductivity tensor \cite{CALDERIN2017118}. 

Using Eq.~(\ref{eq:di_f}) and Eq.~(\ref{eq:di_f2}), one can express the imaginary part of the dynamic density response function in terms of the dynamic conductivity $\sigma_{\rm KG}(\omega)$ as
\begin{eqnarray}\label{eq:chi_KG}
    \chi^{*}_{\rm KG}(\omega)\underset{\vec q\to0}{=\joinrel=}v(q)\chi_{\rm KG}(\vec q, \omega)\equiv{\rm Im}\left[\frac{1}{1+\frac{4\pi i}{\omega}\sigma_{\rm KG}(\omega)} \right],
\end{eqnarray}
where we introduced the dynamic density response function scaled by $v(q) ={4\pi}/{q^2}$. We note that $\chi^{*}_{\rm KG}(\omega)$ does not depend on $\vec q$ because the electron density response exhibits a quadratic dependence $\propto q^2$ in the small wavenumber limit for a given fixed configuration of ions \cite{Arista_pra_1984, Moldabekov_pop_2018, Moldabekov_Electronic_Structure_2025, Moldabekov_prb_2023}.

From Eq.~(\ref{eq:chi_KG}) and Eq.~(\ref{eq:fluc_dis}), we find the scaled DSF in the Kubo-Greenwood approximation:
\begin{eqnarray}
        S^{*}_{\rm KG}(\omega)\underset{\vec q\to0}{=\joinrel=}v(q)S_{\rm KG}(\vec q,\omega)\equiv -\frac{\hbar^2}{n}~\frac{1}{1-e^{-\hbar \omega/k_BT}}~{\rm Im}\left[\chi^{*}_{\rm KG}(\omega) \right].
\end{eqnarray}

Since the ITCF calculation involves only the frequency integration, taking the limit $\vec q\to0$ in the DSF explicitly is not required. Instead, $S^{*}_{\rm KG}(\omega)$ can be used directly to perform the $\eta$ convergence test in the imaginary time domain using
\begin{equation}
    F^{*}_{\rm KG}(\tau)\underset{\vec q\to0}{=\joinrel=}v(q)F_{\rm KG}(\tau)\equiv \int_{-\infty}^\infty \textnormal{d}\omega\ S^{*}_{\rm KG}(\omega)\ e^{-\tau\omega}.
\end{equation}

After identifying the $\eta$-convergence range, filtering of narrow-band
fluctuations can be applied to $S^{*}_{\rm KG}(\omega)$. Following this, the imaginary part of the scaled density response function $\chi^{*}_{\rm KG}(\omega)$ filtered from narrow-band fluctuations can be computed. The real part of the response function $\chi^{*}_{\rm KG}(\omega)$  filtered from narrow-band fluctuations is derived using the Kramers-Kronig relations. 

Finally, the data with attenuated narrow-band fluctuations for the dynamic dielectric function and conductivity in the Kubo-Greenwood approximation can be obtained using the relations $\varepsilon_{\rm KG}^{-1}(\omega) = 1 + \chi^{*}_{\rm KG}(\omega)$ and $\sigma_{\rm KG}(\omega) = \frac{i\omega}{4\pi} \left[1 - \varepsilon_{\rm KG}(\omega)\right]$.

\begin{figure*}[t!]
	\centering
    \begin{subfigure}{}
		\includegraphics[width=0.8\linewidth]{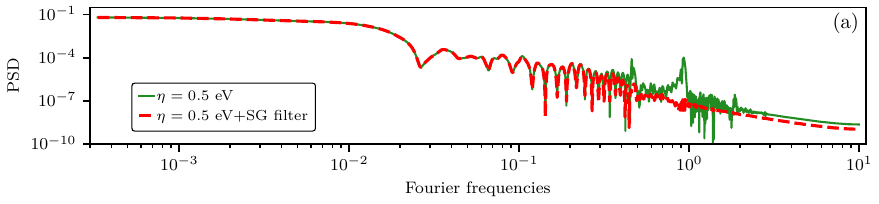}
		\label{fig:subfigA}
	\end{subfigure}
    
    \begin{subfigure}{}
		\includegraphics[width=0.8\linewidth]{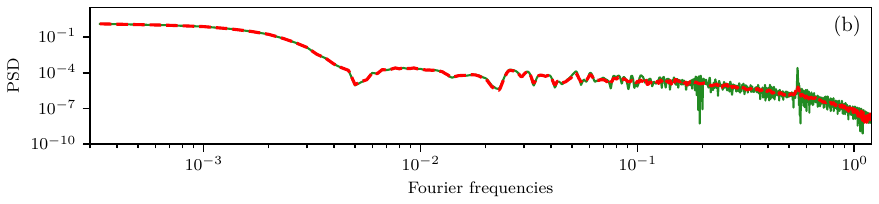}
	\end{subfigure}
\caption{Power spectral density of the DSF of aluminum at $T = 25~\text{eV}$ for (a) $q= 1.0~\textup{~\AA}^{-1}$ and (b) $q = 7.75~\textup{~\AA}^{-1}$. The reference data with $\eta = 0.5~\text{eV}$  show noticeable narrow-band fluctuations at high Fourier frequencies, which are significantly reduced through the application of constraint-based filtering.}
	\label{fig:PSD_Al}
\end{figure*}

\subsection{The SG filter parameters used for isochorically heated Al}\label{s:filt_par}


 
For the DSF of isochorically heated Al ($T=25~{\rm eV}$) at $q= 1.0~\textup{~\AA}^{-1}$, we used the third-order polynomial based SG filter with a $w=3.65~{\rm eV}$ window for the data at $\omega\leq 50~{\rm eV}$. At $\omega> 50~{\rm eV}$, we used the mean of the results generated by applying the window parameters $w=3.75~{\rm eV}$ and $w=3.5~{\rm eV}$.
At $q= 7.75~\textup{~\AA}^{-1}$ with $10\times10\times10$ k-point grid, 
for the DSF values at $\omega>50~{\rm eV}$, we used the first-order polynomial-based SG filter with the window $w=8.55~{\rm eV}$.  For the used filter parameters, we have an efficient attenuation of the narrow-band fluctuations (as shown in Fig.~\ref{fig:PSD_Al} for the data with $10\times10\times10$ k-point grid) and a negligible impact on the ITCF, with both  $ \bar{F}_n(q)/\bar{F}(q, \beta/2) $ and  $ \sigma_n(q)/\bar{F}(q, \beta/2) $ being $<10^{-3} [\%]$.
For the data computed using 
$18\times18\times18$ k-point grid, we used the first-order polynomial-based SG filter with $w=4.05~{\rm eV}$ at frequencies outside of the range corresponding to 2s-2p feature.

\section*{Funding} 
This work was partially supported by the Center for Advanced Systems Understanding (CASUS) which is financed by Germany’s Federal Ministry of Education and Research (BMBF) and by the Saxon state government out of the State budget approved by the Saxon State Parliament.
This work has received funding from the European Research Council (ERC) under the European Union’s Horizon 2022 research and innovation programme
(Grant agreement No. 101076233, "PREXTREME"), and from the European Union's Just Transition Fund (JTF) within the project \emph{R\"ontgenlaser-Optimierung der Laserfusion} (ROLF), contract number 5086999001, co-financed by the Saxon state government out of the State budget approved by the Saxon State Parliament.
The PIMC calculations were carried out at the Norddeutscher Verbund f\"ur Hoch- und H\"ochstleistungsrechnen (HLRN) under grant mvp00024 and on a Bull Cluster at the Center for Information Services and High Performance Computing (ZIH) at Technische Universit\"at Dresden. 
MP is funded by a grant from the US Department of Energy, No.~DE-SC0024496.

\section*{Competing interest}
The authors have no competing interests that might be perceived to influence the results and/or discussion reported in this paper.

\section*{Author Contributions}
ZM: Conceptualization (lead); Data curation (lead); Formal analysis (lead); Investigation (lead); Methodology (lead); Validation (lead); Visualization (lead); Writing – original draft (lead).
TD: Conceptualization (equal); Data curation (equal); Formal analysis (equal); Funding acquisition (lead); Methodology (equal);  Validation (equal); Writing – original draft (equal).
SS, UHA, JV, and MP  contributed to data generation, manuscript drafting, and critical discussions. All authors reviewed the manuscript.

\section*{Data Availability}
The data supporting the findings of this study are available on the Rossendorf Data Repository (RODARE)~\cite{moldabekov_zhandos_2026_4513}.


\bibliography{sn-bibliography}



\end{document}